\UseRawInputEncoding 
\documentclass[twocolumn,superscriptaddress,nobibnotes,nofootinbib,floatfix,aps,prl,citeautoscript,reprint,longbibliography]{revtex4-1}

\usepackage{graphicx}
\usepackage{epsfig}
\usepackage{subfigure}
\usepackage{color}
\usepackage{latexsym}
\usepackage{amsmath,bm,multirow}
\usepackage{amsfonts}
\usepackage{dcolumn}           
\usepackage{natbib}
\usepackage{physics}

\usepackage{ragged2e}

\usepackage[draft]{changes} 
\usepackage[normalem]{ulem}

\newcommand{\nw}{Department of Materials Science and Engineering, Northwestern University, Evanston, IL 60208, USA}

\usepackage[draft]{changes} 
\usepackage[normalem]{ulem}
\usepackage{soul}

\begin{document}

\title{Microscopic mechanism of unusual lattice thermal transport in TlInTe$_2$}

\author{Koushik Pal}
\email{koushik.pal.physics@gmail.com}
\affiliation{\nw}

\author{Yi Xia}
\affiliation{\nw}

\author{Chris Wolverton}
\email{c-wolverton@northwestern.edu}
\affiliation{\nw}



\begin{abstract}
We investigate the microscopic mechanism of  ultralow lattice thermal conductivity ($\kappa_l$) of TlInTe$_2$ and its weak temperature dependence using a unified theory of lattice heat transport that considers contributions arising from the particle-like propagation  as well as wave-like tunneling  of phonons. While we use the Peierls-Boltzmann transport equation (PBTE) to calculate the particle-like contributions ($\kappa_l$(PBTE)), we explicitly calculate the off-diagonal  (OD) components of the heat-flux operator within a first-principles density functional theory framework to determine the contributions ($\kappa_l$(OD)) arising from the wave-like tunneling of phonons. At each temperature, T, we anharmonically renormalize the phonon frequencies using the self-consistent phonon theory including quartic anharmonicity, and utilize them  to calculate $\kappa_l$(PBTE) and $\kappa_l$(OD).  With the combined inclusion of $\kappa_l$(PBTE), $\kappa_l$(OD), and additional grain-boundary scatterings, our calculations successfully reproduce the experimental results. Our analysis shows that large quartic anharmonicity of TlInTe$_2$ (a) strongly hardens the low-energy phonon branches, (b) diminishes the three-phonon scattering processes at finite T, and (c) recovers the weaker than  T$^{-1}$ decay of the measured $\kappa_l$.
\end{abstract}

\maketitle

\section{Introduction} 
Crystalline semiconductors with ultralow lattice thermal conductivity ($\kappa_l$) are important  for effective  utilization 
and management of thermal energy in high-performance  thermoelectrics \cite{bell2008cooling, snyder2008complex, cahill2014nanoscale, zhou2018routes},  photovoltaics \cite{green2017energy, xie2020all, lee2017ultralow, acharyya2020intrinsically}, thermal barrier coatings \cite{darolia2013thermal}, and  thermal data storage  devices \cite{matsunaga2011phase}. Although  it is  quite  natural for compounds with  complex crystal structures having large unit cells \cite{li2016low, shi2018new} and heavy atoms to exhibit   low-$\kappa_l$, relatively simple crystalline materials having small unit cells and even with light  atoms \cite{peng2018unlikely} could  also possess ultralow-$\kappa_l$ due to the presence of rattler atoms  \cite{christensen2008avoided}, strong   lattice anharmonicity \cite{li2015orbitally,nielsen2013lone, zhang2012first, pal2019intrinsically, skelton2016anharmonicity},  or  bonding  heterogeneity  \cite{qiu2014part, zhou2017promising, pal2018bonding, samanta2020intrinsically}. Investigation of the  microscopic mechanism behind  the  ultralow-$\kappa_l$  that often approaches the glassy limit in ordered  compounds is not only fundamentally   interesting,  but it  also helps to  unravel the complex  correlation  between the crystal structure, bonding, and anharmonic   lattice dynamics. Results of such investigations provide new  criteria to find   hitherto unknown low-$\kappa_l$ materials as well as paves the way to  engineer the   heat transport properties in  already known compounds. 

Despite significant research efforts, a comprehensive theoretical  understanding of  the mechanism behind  extremely poor heat transport  in ultralow-$\kappa_l$ materials,  which approach  the   limit of their theoretical    minimum  ($\kappa_{min}$) has remained challenging \cite{cahill1992lower,  allen1993thermal, allen1989thermal}. This is in  part due to the fact that many of these materials  are often so highly anharmonic that a harmonic description of the phonon frequencies fails to describe  the lattice dynamics of the compounds correctly.   Hence,  a finite temperature  treatment  of the  phonon modes  becomes necessary to  account for the  renormalization of the phonon frequencies arising from the temperature induced anharmonic effects.  In some cases,  the mean  free paths of the phonon modes approach the smallest atomic  distance in the crystal, leading to a breakdown of  the  conventional  particle-like description of phonons towards a glass-like thermal conductivity \cite{sun2010lattice}.  However, recent   theoretical and computational developments \cite{zhou2014lattice, tadano2015self, tadano2018quartic, errea2011anharmonic, simoncelli2019unified, isaeva2019modeling} have enabled the  treatment  of   phonons at finite  temperature considering   anharmonicity arising from high-order phonon-phonon interactions,  and examination  beyond  the particle-like description  by including the contributions  arising from the  wave-like tunneling of phonons  \cite{simoncelli2019unified, isaeva2019modeling}.  In recent theoretical studies, it was shown that the ultralow $\kappa_l$ in many well known family of crystalline solids such as clathrates \cite{tadano2018quartic}, double-halide perovskites \cite{klarbring2020anharmonicity}, tetrahedrites \cite{xia2020microscopic}, Tl$_3$VSe$_4$ \cite{xia2020particlelike} can only be explained successfully if higher-order anharmonic phonon-phonon interactions are taken into account in the description of their lattice dynamics and phonon transport \cite{errea2015high, tadano2015self}.

\begin{figure*}
\centering
\includegraphics[height=5.5cm, width=13cm, trim={0cm  2cm 0cm 3cm}, clip]{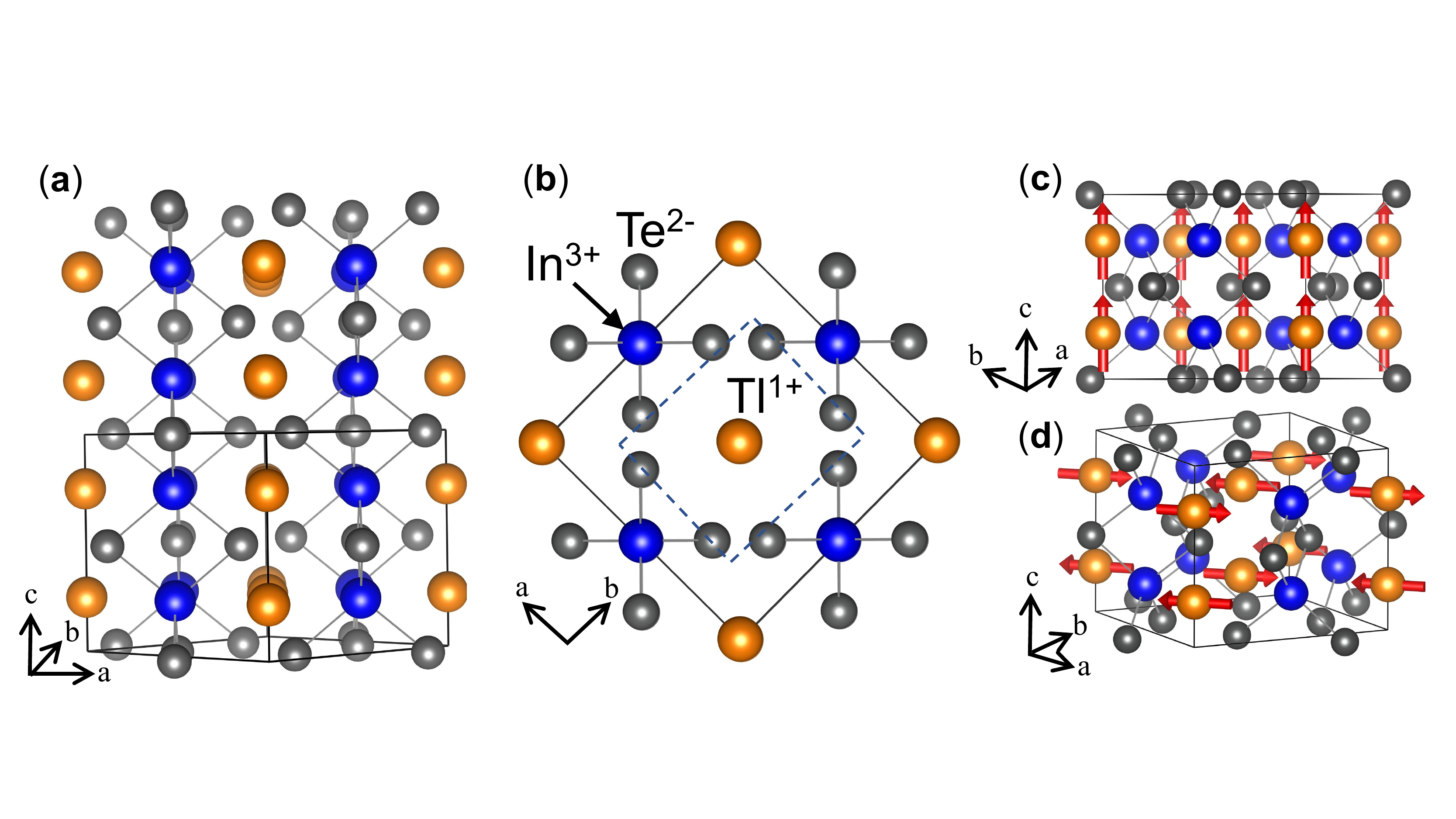}
\caption{Crystal structure, bonding, and rattling vibrations. (a) Chain-like crystal structure of TlInTe$_2$, where In$^{3+}$ cations (blue sphere) are tetrahedrally coordinated by Te$^{2-}$ anions (dark grey sphere), forming InTe$_4$ tetrahedron. These InTe$_4$ tetrahedra share their edges, and extend along 
the c-axis, giving rise to a chain-like structure. Tl$^{1+}$ cations (orange sphere) occupy the spaces between the 
chains and stabilize the structure through electron transfer from Tl$^{1+}$ to the [InTe$_2$]$^{1-}$ sublattices. The conventional unit cell is outlined by the solid black line. 
(b) A top-down view of the crystal structure, where it can be seen that  Tl atom occupies the empty space inside the 
square anti-prismatic (outlined by dashed blue line) environment  formed by  eight Te atoms.   Visualization of the 
(c)  longitudinal (R$_{LO}$)  and (d) transverse (R$_{TO}$) optical rattling (R) phonon modes at the X point in the 
Brillouin zone, where the atomic displacements are indicated by the red arrows.}
\end{figure*}

In this work, we develop a  microscopic understanding and uncover the  physical principles underlying the unusual lattice thermal transport properties of TlInTe$_2$ which exhibits  ultralow-$\kappa_l$ that is close to its theoretical minimum \cite{matsumoto2008systematic, jana2017intrinsic} and shows weak (milder than T$^{-1}$ decay) temperature dependence in $\kappa_l$. TlInTe$_2$   represents a class of structurally similar  ABX$_2$ (A=Tl$^{1+}$, In$^{1+}$; B=Tl$^{3+}$, In$^{3+}$, Ga$^{3+}$; X=Se$^{2-}$, Te$^{2-}$) compounds, many of which are shown to exhibit ultralow-$\kappa_l$  due to  the presence of rattler cation at the A-site \cite{jana2016origin, matsumoto2008systematic, jana2017intrinsic, dutta2019ultralow}.  The  rattler cations manifest nearly dispersion-less optical phonon branches at  low-energies which  effectively scatter the heat-carrying phonons by creating additional phonon scattering channels  \cite{tadano2015impact, li2015ultralow, pal2019intrinsically}. Since phonon scattering rates crucially depend on the phonon frequencies \cite{li2014shengbte} and the rattling phonon  modes are  quite sensitive to temperature, an  accurate treatment of the thermal transport properties in the above compounds requires the treatment of  phonons at finite  temperature including anharmonic effects \cite{tadano2015self, van2016high}. However, a comprehensive theoretical understanding of the lattice thermal transport   including the effects of temperature-induced anharmonic  renormalizations to the phonon frequencies is missing in this family of compounds.  Here,  we use a unified theory of lattice thermal conductivity that considers the contributions arising from the particle-like propagation as well as wave-like tunneling of phonons. We utilize the  phonon frequencies that are renormalized at finite temperatures including the effects of quartic anharmonicity to calculate both these contributions to $\kappa_l$.

\begin{figure*}
 \centering
 \includegraphics[height=7cm, width=18cm, trim={0cm  4cm 0cm 2cm}, clip]{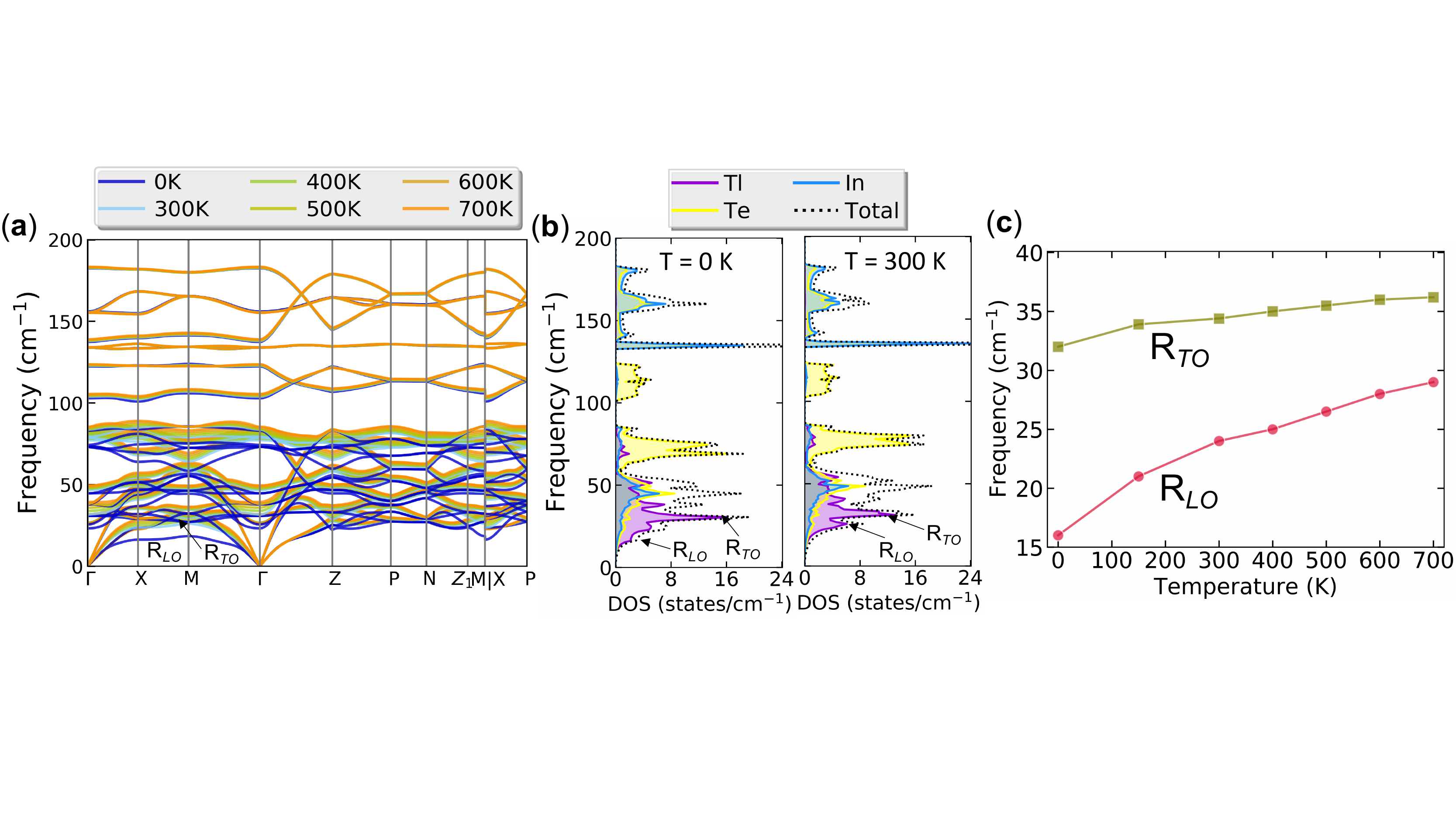}
 \caption{Temperature-induced anharmonic phonon renormalization. (a) Anharmonically renormalized  phonon dispersions  of TlInTe$_2$ at finite temperatures (0-700 K).  The longitudinal optical  (R$_{LO}$) and transverse optical (R$_{TO}$) rattling phonon branches at T = 0 K are marked with arrows. (b) Atom-resolved   phonon density of states of TlInTe$_2$ at 0 K and 300 K, where the peaks associated with the R$_{LO}$  and R$_{TO}$ phonons  are indicated. (c) Variations of the R$_{LO}$ and R$_{TO}$ modes at the X-point in the Brillouin zone as a function of temperature.}
 \end{figure*}

The particle-like contributions  ($\kappa_l$(PBTE)) which are obtained after solving the Peierls-Boltzmann transport equation (PBTE) only account for the diagonal components of the heat-flux operator \cite{hardy1963energy,simoncelli2019unified,allen1999diffusons, lv2016non}:  $J^{\alpha} = -\kappa_l^{\alpha\beta}\nabla T^{\beta}$, where $\nabla T$ is the temperature gradient  and $\alpha$, $\beta$ are the Cartesian coordinates.  We  also calculate the  off-diagonal (OD) contributions, $\kappa_l$(OD),  associated with  the  OD components of the heat-flux operator, which are not present in the PBTE formalism.  $\kappa_l$(OD) is related to the wave-like tunneling  of phonons, which is the heat carried  by the  coupled vibrational  eigenstates as a result of the loss of coherence  between them  \cite{hardy1963energy, kane2012zener, simoncelli2019unified, isaeva2019modeling}.  In a recent theoretical work, Simoncelli et  el. \cite{simoncelli2019unified} combined both the pictures (i.e.,  particle-like  and  wave-like) of phonon transport in a unified theory   of lattice heat  transport,  where  the total $\kappa_l$  is given by $\kappa_l^{\alpha\beta}(\text{tot})  =  \kappa_l^{\alpha\beta}(\text{PBTE}) + \kappa_l ^{\alpha\beta}(\text{OD})$, which successfully describes the  $\kappa_l$ of  anharmonic crystals, harmonic glasses as well as complex compounds such as  tetrahedrites  \cite{simoncelli2019unified, xia2020particlelike, xia2020microscopic}.  Our  calculated  $\kappa_l$ within the PBTE formalism in conjunction with the OD contributions using the renormalized phonon frequencies, and additional grain-boundary scatterings  successfully reproduce two sets of  available experimental results \cite{matsumoto2008systematic,jana2017intrinsic} of TlInTe$_2$.  Our analysis reveals  that TlInTe$_2$ possesses large quartic  anharmonicity that (a) strongly  hardens  the  low-energy rattling and other optical phonon branches with temperature, (b)  diminishes the    three-phonon  scattering  rates at finite temperatures, and  (c) recovers  the magnitude as well as the correct  T-dependence of  $\kappa_l$ that shows weaker than   T$^{-1}$  decay found in experimental measurements.

\section{Results}
\subsection{Crystal structure and anharmonicity}
TlInTe$_2$ has a chain-like  body-centered tetragonal (space group: I4/mcm) crystal structure (Fig. 1a) with eight atoms in the primitive unit cell. Within the unit cell,  In$^{3+}$ cations are covalently bonded to four Te  atoms,  forming the   InTe$_4$ tetrahedra which share their edges and extend like  chains along the crystallographic  c-axis.  The empty  spaces between these chains are bridged by the Tl atoms which stabilize the structure  through electron transfer from  Tl$^{1+}$ cation to the [InTe$_2$]$^{-1}$ anion sublattices. Analysis of the second-order interatomic force constants (IFCs)    reveals that In$^{3+}$ are strongly bonded to the lattice, whereas the Tl $^{1+}$ cations are only loosely connected (see  Supplementary Figure 2). In the crystal structure, Tl atoms are coordinated by eight Te atoms in  a square  anti-prismatic  coordination    environment (Fig. 2b) forming an oversized cage (also known as Thompson cube)  inside which the Tl atoms rattle due  to their weaker chemical bonding. The rattling (R) vibrations  have two components: (a) the longitudinal optical (R$_{LO}$) and (b) transverse optical  (R$_{TO}$) branches, which  are shown in Fig. 1c and Fig. 1d,   respectively.   As  will be  discussed later, the  R$_{LO}$ branch  is strongly affected by the temperature, which suppresses the   phonon-scattering  rates at high  temperature, giving rise to a  milder  T$^{-0.86}$ decay of the calculated  $\kappa_l$ that is   closer to the   temperature  dependence of  T$^{-0.88}$ and T$^{-0.82}$  found in  the  experimentally  measured $\kappa_l$ of  TlInTe$_2$  in   two sets of  experiments \cite{jana2017intrinsic,matsumoto2008systematic}.

Since the presence of strong anharmonicity is an important characteristic of many ultralow-$\kappa_l$ compounds,  it is necessary to asses its strength in TlInTe$_2$. Anharmonicity of the phonon modes is estimated by the mode Gruniesen parameters  ($\gamma_{\lambda} = -\frac{d ln\omega_{\lambda}}{d lnV}$)  that quantify the  change of the  phonon frequencies ($\omega_{\lambda}$) with   respect to  the change in the unit cell volume  ($V$), where $\lambda$ is a composite index that combines the wave vector ($q$) and phonon branch index ($s$).  While for weakly anharmonic solids $\gamma_{\lambda}$'s are close to 1,  for highly anharmonic materials $\gamma_{\lambda}$'s become much larger than 1. Some examples of compounds that possess large    $\gamma_{\lambda}$'s (hence, strong anharmonicity) and ultralow-$\kappa_l$ are  AgBiSe$_2$  \cite{nielsen2013lone}  and SnSe \cite{zhao2014ultralow,li2015orbitally}.  Previous studies \cite{jana2016origin,jana2017intrinsic,dutta2019ultralow,wu2019unusual}   have shown that the phonon modes of this ABX$_2$ family of materials  exhibit  large  $\gamma_{\lambda}$'s. Thus, in the  presence of strong anharmonicity, the harmonic description of the phonon frequencies of the compounds in this family becomes inappropriate at finite temperatures as the anharmonicity induces multi-phonon interactions, giving rise to shifts in the phonon frequencies and  broadening  of the phonon states.  As a consequence, the phonon frequencies of these anharmonic solids are expected   to have strong renormalization effects at finite temperatures, which can crucially  alter both the magnitude and T-dependence of the  $\kappa_l$ that deviates from the ideal T$^{-1}$ behavior found in weakly  anharmonic solids.  Indeed, the weak temperature dependence of the experimentally measured \cite{jana2017intrinsic,matsumoto2008systematic} $\kappa_l$ of TlInTe$_2$, which decays as   T$^{-0.88}$ and  T$^{-0.82}$, signifies the  prevalence of severe anharmonicity in the crystal structure  of TlInTe$_2$.

\begin{figure*}
 \centering
 \includegraphics[height=10cm, width=18cm, trim={1cm  0cm 0cm 0cm}, clip]{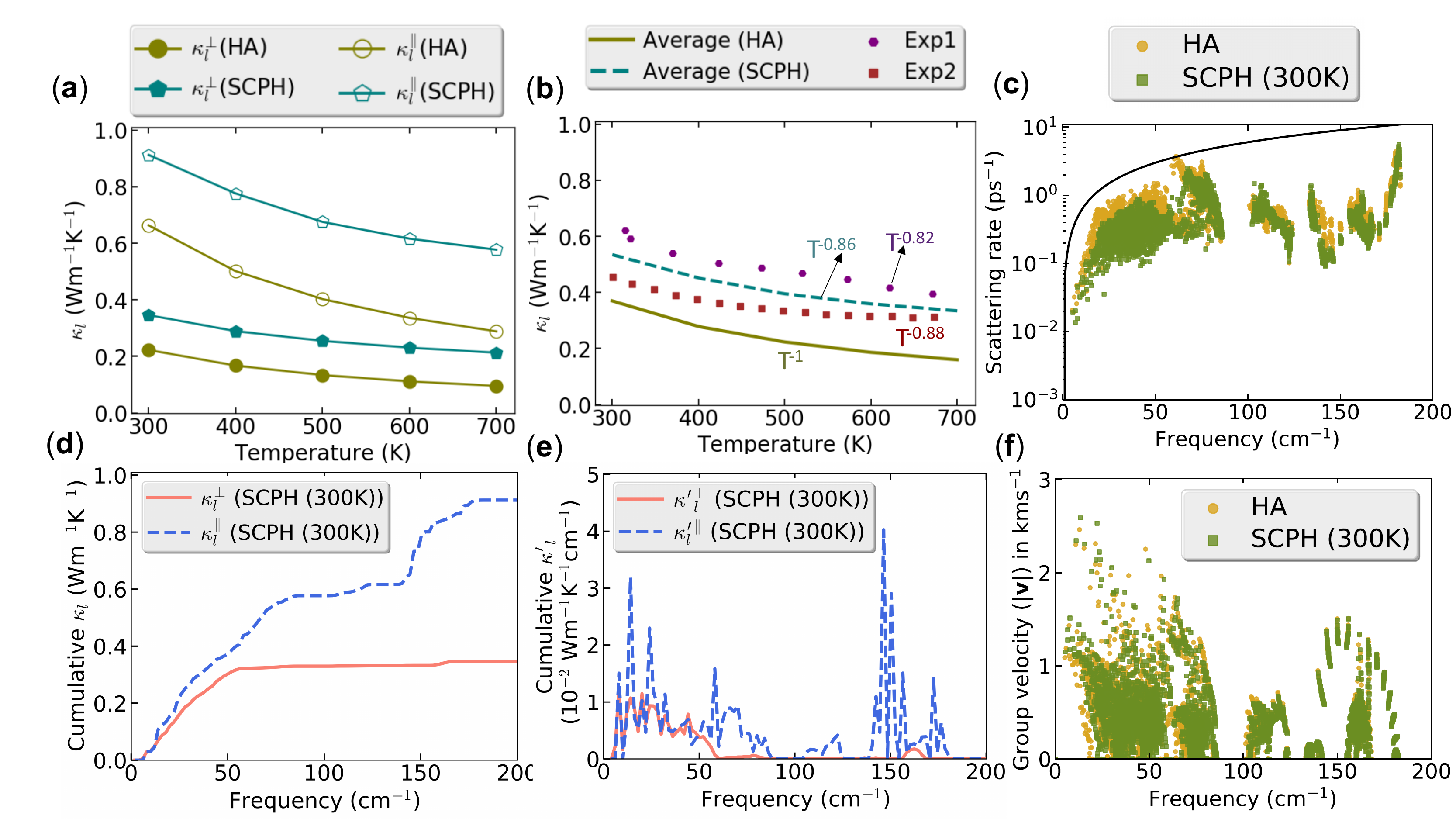}
 \caption{Heat transport due to particle-like propagation of phonons. (a) Particle-like contributions to the lattice thermal conductivity ($\kappa_l$) of TlInTe$_2$ calculated
 utilizing  the harmonic (i.e., HA)  and   the anharmonically renormalized  (i.e., SCPH) phonon frequencies using the PBTE.
 $\kappa_l^{\Vert}$ and $\kappa_l^{\bot}$ are the  components of $\kappa_l$, which are  parallel and 
 perpendicular to the chain directions in the crystal structure of TlInTe$_2$, respectively.
 (b) The averages of the $\kappa_l^{\Vert}$ and $\kappa_l^{\bot}$  components are compared against
 the experimentally measured values which are denoted with Exp1 and Exp2 taken from 
 Ref. \cite{matsumoto2008systematic} and Ref. \cite{jana2017intrinsic}, respectively. (c) Three-phonon scattering rates,
 and (f) phonon group velocities rates obtained using the harmonic (i.e., HA)  and anharmonically renormalized 
 (i.e., SCPH) phonon frequencies at T = 300 K.  Mode contributions to  $\kappa_l^{\Vert}$ and $\kappa_l^{\bot}$  are analyzed
 through (d) the cumulative plots   and (e) their first-order derivatives with respect to the  anharmonically renormalized phonon frequency at T = 300 K.  The solid black line in (c) assumes the  scattering rates of the phonon modes to be twice their frequencies according to the Cahill-Watson-Pohl model \cite{cahill1992lower}.}
 \label{fig:anakappa}
 \end{figure*}

\subsection{Temperature-induced anharmonic phonon renormalization}
We use the self-consistent phonon theory (SCPH) \cite{cowley1968anharmonic, werthamer1970self, errea2011anharmonic, tadano2015self} to renormalize the  phonon frequencies of  TlInTe$_2$ at finite  temperatures  including  anharmonic effects \cite{zhou2014lattice} which are treated as the phonon self-energies \cite{klein1972rise, errea2011anharmonic}. Within the SCPH theory,  the anharmonically renormalized phonon frequency is determined from the pole of the many-body  Green's   function.  Considering only the first-order  contribution to the phonon self-energy arising from the quartic anharmonicity, the SCPH equation \cite{tadano2015self} is written as   $\Omega^2_{\lambda} = \omega^2_{\lambda}+2\Omega_{\lambda}I_{\lambda}$, where $\omega_{\lambda}$ is the  harmonic phonon frequency at  T = 0 K  and $\Omega_{\lambda}$ is the anharmonically renormalized phonon frequency at finite T.  The quantity $I_{\lambda}$  is defined as: $I_{\lambda} = \frac{\hbar}{8N} \sum_{\lambda_{1}} \frac{V^{(4)}(\lambda,-\lambda,\lambda_{1},-\lambda_{1})}{\Omega_{\lambda}\Omega_{\lambda_{1}}} \left[ 1+2n\left(\Omega_{\lambda_{1}}\right) \right]$, where $N$, $\hbar$, $n$, and $V^{(4)}(\lambda,-\lambda,\lambda_{1},-\lambda_{1})$ are the number of sampled wave vectors, the reduced Planck constant, the temperature-dependent phonon population, and the reciprocal representation of the fourth-order IFCs, respectively. The temperature dependence of the SCPH equation is contained in the phonon population term that obeys the Bose-Einstein statistics. Since $\Omega_{\lambda}$ and  $I_{\lambda}$ are inter-dependent on each other, the SCPH equation is solved self-consistently until  the convergence in $\Omega_{\lambda}$ is achieved.

We present the anharmonically renormalized phonon dispersions of TlInTe$_2$  in Fig. 2a in the temperature range between 0-700 K.  The  harmonic phonon dispersion (i.e., T= 0 K) of TlInTe$_2$ exhibits two groups of  low-frequency  optical phonon branches with very small dispersions, that are characteristic of the rattler atoms in the crystal structure. The lowest-energy rattling  phonon branch (denoted as R$_{LO}$) arises from the longitudinal  vibration of   Tl$^{1+}$ ions along the direction of the InTe$_4$ chains (Fig. 1c) inside the hollow Thompson cube. The second group of  rattling phonon branches (R$_{TO}$) appears above R$_{LO}$,  where the Tl$^{1+}$  cations vibrate along the transverse  direction  (i.e., perpendicular to the InTe$_4$ chains, Fig. 1d).  Both R$_{LO}$  and R$_{TO}$ phonons are highly localized at T = 0 K as revealed by analysis based on the phonon participation ratio \cite{pailhes2014localization, tadano2015impact}, which does not change as they are anharmonically renormalized at finite temperatures (see see Supplementary Figure 6 and Supplementary Note 4). The  atom-resolved phonon density of states at T=0 and T = 300 K are shown in Fig. 2b, which clearly show the two peaks associated with R$_{LO}$ and R$_{TO}$ phonons, which gradually convolute as the temperature increases. It is  seen from Fig. 2a that only the phonons up to 100 cm$^{-1}$ are strongly hardened while the phonons above  100 cm$^{-1}$ show very weak hardening or softening.  The most drastic temperature-induced changes are observed in the frequency hardening of R$_{LO}$ and R$_{TO}$ phonons (Fig. 2c). It is interesting to note that the frequency of R$_{LO}$ phonon is smaller than that 
of R$_{TO}$  due to the chain-like crystal structure and weak bonding of Tl atoms which vibrate slowly but with much 
larger  amplitudes  along the c-direction in the empty space within the lattice.

\subsection{Particle-like contributions to $\kappa_l$}
We start our analysis of the thermal conductivity by examining the calculated $\kappa_l$ obtained from solving the PBTE using the harmonic (i.e., T= 0 K)  and  anharmonically renormalized phonon frequencies (at finite T)  obtained using the SCPH method. Hereafter, we denote these results by  $\kappa_l$(HA), and  $\kappa_l$(SCPH), respectively, as shown in Fig.  3a.  It is seen that $\kappa_l$(HA)  changes  significantly when the renormalized phonon frequencies are incorporated.  For example, $\kappa_l^{\bot}$(HA) and  $\kappa_l^{\Vert}$(HA) increase by  55 \% and 38 \%, respectively, at 300 K when the renormalized phonon  frequencies are used to solve the PBTE. Here, $\Vert$ and $\bot$ symbols indicate components of $\kappa_l$ parallel and perpendicular to the chain direction (i.e., the c-axis) in the crystal structure of TlInTe$_2$, respectively.  We compare the average $\kappa_l$ calculated  under both HA  and SCPH methods with that of available experimental measurements \cite{jana2017intrinsic,matsumoto2008systematic} of  $\kappa_l$ in TlInTe$_2$. From  Fig. 3b,  we see that the average    $\kappa_l$(HA) is significantly underestimated  in magnitude compared to the two sets of available experiments \cite{jana2017intrinsic,matsumoto2008systematic}.  While  $\kappa_l$(HA)  decays as   T$^{-1}$ according to  well-known behavior of the lattice thermal conductivity  found in weakly anharmonic solids, the two  experimentally  measured $\kappa_l$ data decay as T$^{-0.82}$ and   T$^{-0.88}$. The deviation from T$^{-1}$  behavior and the weaker temperature dependence  signify the presence of a strong higher-order anharmonicity in  TlInTe$_2$, and thus necessitate  the anharmonic renormalization of the harmonic phonon  frequencies. The calculated average of  $\kappa_l$ within the SCPH method  increases in magnitude with respect to $\kappa_l$(HA),  but still does not agree well with either  sets of experimentally measured values of $\kappa_l$.  However, the effect of  anharmonic   renormalization   significantly improves the   temperature dependence, which varies as T$^{-0.86}$  (Fig. 3b), bringing it closer to the experimental observations.
    
\begin{figure*}
 \centering
 \includegraphics[height=11cm, width=19cm, trim={2.0cm  0cm 0cm 0cm}, clip]{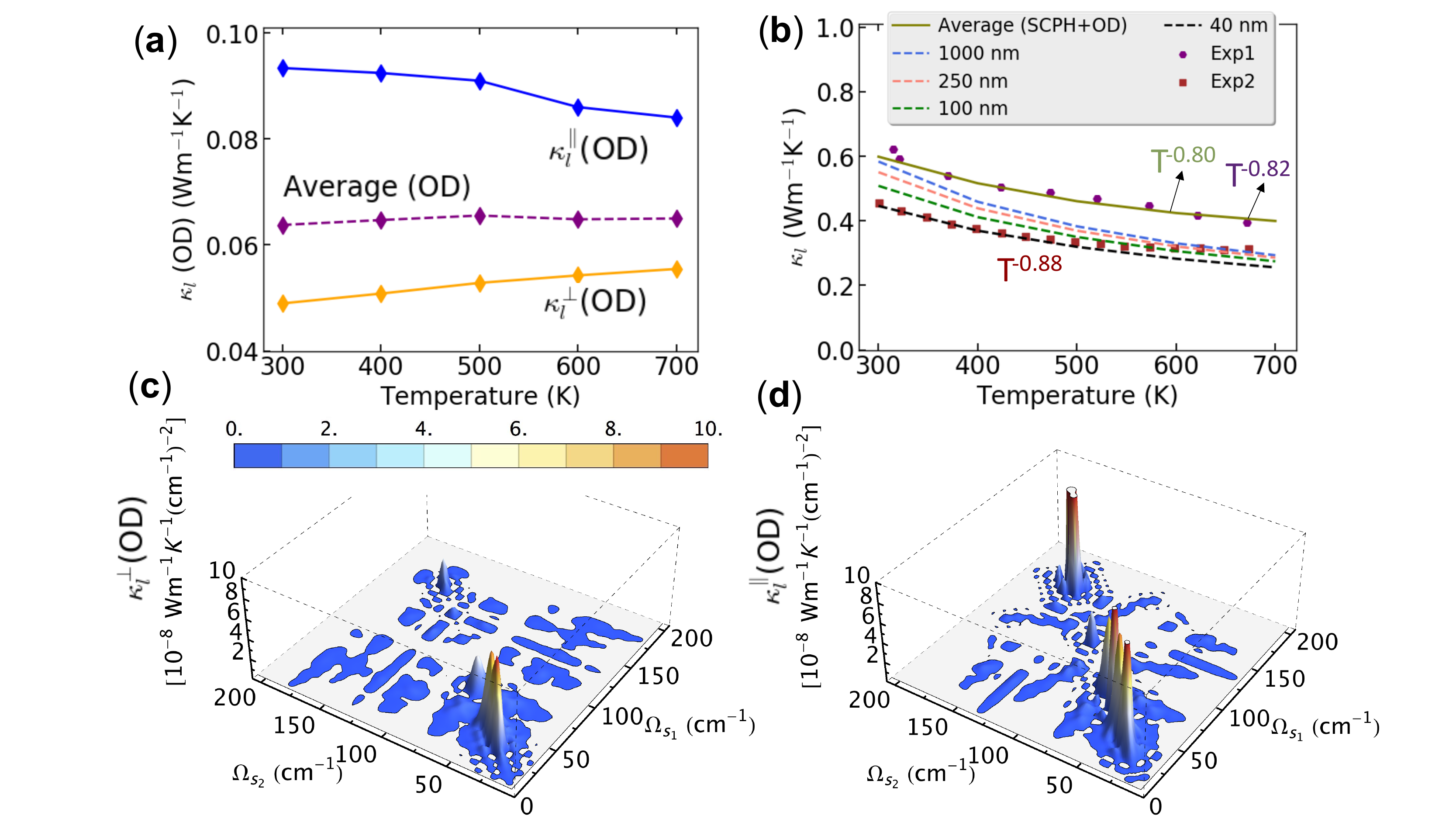}
 \caption{Heat transport due to wave-like tunneling of phonons. (a) The  off-diagonal (OD) contributions to $\kappa_l$  (i.e., $\kappa_l^{\Vert}$,  $\kappa_l^{\bot}$, and 
 their average) calculated using the anharmonically renormalized  phonon frequencies at each temperature. 
 (b) Total $\kappa_l$ (calculated within SCPH method) including the OD contributions reproduces the Exp1 \cite{matsumoto2008systematic} well. Additional grain-boundary scatterings on top of SCPH+OD contributions to $\kappa_l$ reproduce Exp2 \cite{jana2017intrinsic}. Grain sizes ranging from 40 to 1000 nm are considered here.
Three-dimensional visualizations of the mode specific contributions to $\kappa_l$(OD) as a function of 
anharmonically renormalized   phonon frequencies (i.e., $\Omega_{s}$)  at T = 300 K for the 
(c)   $\kappa_l^{\bot}$ and (d) $\kappa_l^{\Vert}$ components.  These quantities are plotted per cross-sectional 
area of the Brillouin zone to clearly highlight the phonon frequencies that primarily  contribute to the
$\kappa_l$(OD). A color scale corresponding to the variation of colors in Figs. 4c and 4d is shown, which has the same unit as indicated in the z-axes of both figures.}
 \label{fig:anakappa}
 \end{figure*}

Next, we analyze how the anharmonic renormalization affects the two key ingredients that enter into the PBTE (see  Supplementary Note 2),  namely the three-phonon  scattering rates  (Fig. 3c) and the phonon group velocities (Fig. 3f). According to the phonon gas  model, where the phonons behave like particles, the maximum phonon scattering rate of a phonon mode is assumed to be twice its frequency \cite{cahill1992lower}, which is denoted with a black line in Fig. 3c. The HA scattering rates are strongly suppressed by the anharmonic effects in the frequency region below 100 cm$^{-1}$ 
 which enhances the phonon  lifetimes, leading to a large increase in $\kappa_l$(SCPH) compared to $\kappa_l$(HA).
The SCPH scattering rates are well below the solid black line, indicating the dominant particle-like nature of the phonons in TlInTe$_2$ that rules out the existence of  a hopping channel of instantaneously localized vibrations as shown in Ref. \cite{wu2019unusual}. The phonon group velocities are weakly hardened below 100 cm$^{-1}$ (Fig. 3f) due to the anharmonic effects at finite temperatures and its effect on $\kappa_l$(SCPH) is less significant than that of the scattering rates. To examine the phonon mode  specific contributions to $\kappa_l$, we show the cumulative plot of $\kappa_l$(SCPH), and its derivative $\kappa'_l$(SCPH) with respect to renormalized phonon frequency in Fig. 3d and Fig. 3e, respectively. The cumulative plot for    $\kappa_l^{\bot}$  changes rapidly up to  60 cm$^{-1}$ and reaches a plateau above it, showing that only the  acoustic and low-energy optical phonons mainly contribute to it. This is also clear from Fig. 3e, where  large peaks are present mainly up to 60 cm$^{-1}$.   On the other hand, examining $\kappa_l^{\Vert}$  in both Fig. 3e and Fig. 3f, reveals that $\kappa_l^{\Vert}$ has large contributions coming not only from the acoustic and low-energy optical phonons ($<$ 80 cm$^{-1}$) but also from the high-energy optical phonons ($\sim$ 140-180 cm$^{-1}$).

\subsection{Off-diagonal contributions to $\kappa_l$}
To  understand the difference between the measured $\kappa_l$ and calculated $\kappa_l$ (within the SCPH method) of TlInTe$_2$,  we recognize that $\kappa_l$ obtained after solving the PBTE i.e., $\kappa_l$(PBTE) only accounts for  the diagonal components of the heat-flux  operator \cite{simoncelli2019unified,allen1999diffusons, lv2016non}.  We calculate the off-diagonal (OD)  contributions i.e.,  $\kappa_l$(OD)  using the renormalized phonon frequencies and obtain the  total \cite{simoncelli2019unified} $\kappa_l$ as:  $\kappa_l\text{(tot)}  =  \kappa_l(\text{PBTE}) + \kappa_l(\text{OD})$ (see Supplementary Note 1).  It was shown that while  $\kappa_l$(OD) is negligible in compounds like Si and diamond, while it is very important in CsPbBr$_3$ and tetrahedrites \cite{hardy1963energy, simoncelli2019unified, xia2020particlelike, xia2020microscopic}.  One key quantity that enters into the expressions for $\kappa_l$(PBTE) and $\kappa_l$(OD)  is the generalized phonon group velocity operator   (see its expression in SI), $\mathbf{V}_{s_1,s_2}$, where $s_1$ and  $s_2$ are  phonon branch indices.  While $\kappa_l$(PBTE) utilizes only the diagonal components (i.e., $s_1 = s_2$) of   $\mathbf{V}^{s_1,s_2}$,  the  $\kappa_l$(OD) term uses the off-diagonal (i.e., $s_1 \neq s_2$) components of   $\mathbf{V}^{s_1,s_2}$.  Fig. 4a shows the $\kappa_l^{\bot}$ and $\kappa_l^{\Vert}$ components of  $\kappa_l$(OD), and their average as a function of  temperature. Although on an absolute scale these values are  small, for low-$\kappa_l$ compounds, these values are quite significant. For example  at 300K, $\kappa_l^{\bot}$(OD)  and $\kappa_l^{\Vert}$(OD) account for 14 \% and 10 \% contributions to that of $\kappa_l^{\bot}$(SCPH) and  $\kappa_l^{\Vert}$(SCPH)  values, respectively.  However,  with increasing temperature, the relative OD contributions also increase. For example,  the contributions of $\kappa_l^{\bot}$(OD)  and $\kappa_l^{\Vert}$(OD) increase to 26 \% and 15 \%, respectively at 700 K.   When the average $\kappa_l$(OD) is added on top of the average  $\kappa_l$(SCPH) terms, the resulting  $\kappa_l$ agrees very well both in magnitude with one set of experiments, Exp1 \cite{matsumoto2008systematic}   and in temperature  dependence that decays as T$^{-0.80}$ (Fig. 4b).   

\subsection{Effects of grain boundaries}

Having reproduced the first  set of experiments (i.e., Exp1 \cite{matsumoto2008systematic}),  we notice that the measured $\kappa_l$ of TlInTe$_2$ in an another experiment (i.e., Exp2 \cite{jana2017intrinsic}) is lower than the former.  Since the measurements were performed in the polycrystalline samples of the compound in which the presence of  grains  boundaries are generally common, the lower $\kappa_l$ in Exp2 \cite{jana2017intrinsic}  most likely originates from the additional  scatterings of the heat carrying phonons due to the grain boundaries.  We introduced  grain-boundary scatterings   on top of  the three-phonon and isotope scatterings and calculated $\kappa_l$ to see if this additional scattering mechanism can explain the second experiment (i.e., Exp2   \cite{jana2017intrinsic}). Assuming that the boundary scatterings are predominantly diffuse, the grain-boundary scattering rates are given by $\tau^{-1}_{\text{gb},\lambda} = \frac{2\left | \mathbf{v}_{\lambda}\right |}{L}$, where $\mathbf{v}_{\lambda}$ and $L$ are the phonon group velocity and the averaged grain size, respectively. The total  scatterings rates of the phonons are obtained by applying the  Matthiessen's rule,  which are then used to calculate $\kappa_l$  (see Supplementary Note 2 for details).  We  show in Fig. 4b the effect of grain boundary scatterings on $\kappa_l$ as a function of grain size. It can be seen that $\kappa_l$ is significantly modified in magnitudes particularly in low temperature ($<$ 550 K). Our calculations show  that while the phonon-scatterings due to the grains with an average size of 40 nm reproduce Exp2 \cite{jana2017intrinsic}  quite well below T= 550 K, larger grain sizes give rise to better agreement with Exp2 above  550 K (Fig. 4b). 

\section{Discussion}

 After successfully explaining the experimental results, we  now closely examine the mode-specific contributions to the $\kappa_l^{\bot}$(OD) and  $\kappa_l^{\Vert}$(OD), which are shown in Fig. 4c and Fig. 4d,  respectively, as three-dimensional plots averaged over the planar area in the Brillouin zone.  We see that  phonons with very similar frequencies with $s_1 \neq s_2$  (i.e., near the diagonal in Fig. 4c) below 100 cm$^{-1}$  primarily contribute to  $\kappa_l^{\bot}$(OD). On the other hand, quasi-degenerate phonons in the frequency  ranges of  20-100 cm$^{-1}$ and near  180 cm$^{-1}$ mainly contribute to  $\kappa_l^{\Vert}$(OD). It is interesting to note that the anisotropy (i.e., $\kappa_l^{\Vert}$/$\kappa_l^{\bot}$) of the PBTE calculated contributions ($\kappa_l$(SCPH)) is much stronger than the anisotropy present in the $\kappa_l$(OD) contributions.  For example,  $\kappa_l^{\Vert}$/$\kappa_l^{\bot}$  = 2.6 for the PBTE results while for the OD terms,  $\kappa_l^{\Vert}$/$\kappa_l^{\bot}$ =1.8 at T = 300 K. While the anisotropy in the phonon dispersion generally  results in the anisotropy  in $\kappa_l$ obtained from PBTE as the group velocity in the PBTE is related to the single phonon  mode, the much  weaker anisotropy in $\kappa_l$(OD) stems from the fact that the OD heat transport  takes place  between coupled phonon eigenstates where the group velocity is associated  with two different  (quasi degenerate)  frequencies. Thus, $\kappa_l$(OD) has a weak dependence on the slopes of the phonon branches, which partially counteracts the anisotropy in $\kappa_l$(PBTE). It is interesting to see that while $\kappa_l^{\bot}$(OD) increases with temperature,  $\kappa_l^{\Vert}$(OD) decreases with  temperature. Our analysis shows that this opposite temperature-dependence arises from the renormalization of the phonon frequencies at finite temperatures. We show in the  Supplementary Figure 7 that when  $\kappa_l^{\bot}$(OD) and $\kappa_l^{\Vert}$(OD) are calculated using the harmonic (i.e., T= 0 K) phonon frequencies, both the components increase with temperature. However, we note that in all cases, the temperature-dependence of both components of $\kappa_l$(OD) are quite weak.

A previous  attempt  to understand the lattice thermal  conductivity of TlInTe$_2$  was performed by Wu et al. \cite{wu2019unusual}    where the  two-channel model \cite{mukhopadhyay2018two} of thermal  conductivity  was used to explain the  experimentally  measured  $\kappa_l$. The two-channel model has been invoked in cases where the particle-like  description of phonons  becomes insufficient to  describe the observed  $\kappa_l$ of materials. Thus, a  second channel of lattice heat conduction is introduced to  compensate for the $\kappa_l$, which is attributed  to arise  from the localized hopping  among the uncorrelated phonon  oscillators \cite{mukhopadhyay2018two}. In the  two-channel model used by Wu et al. \cite{wu2019unusual},  the  phonon  frequencies were treated within the   harmonic approximation (T= 0 K) with no  effects of temperature, which  were then  used to solve the particle-like  contribution to $\kappa_l$ using the PBTE.  Hence,  the  calculated  \cite{wu2019unusual} $\kappa_l$  underestimated  the  measured $\kappa_l$ values of TlInTe$_2$   \cite{matsumoto2008systematic,jana2017intrinsic}.  To account for  this difference,  the contribution to   $\kappa_l$  coming from the  second channel  was calculated using the  Cahill-Watson-Pohl (CWP)  model \cite{cahill1992lower}  which estimates  the  minimum   $\kappa_l$ achievable in disordered solids.  We note that our calculated $\kappa_l$ within the unified theory that utilizes the anharmonically renormalized phonon frequencies successfully reproduces the experiments without requiring to invoke the second channel of the two-channel  model \cite{mukhopadhyay2018two} of $\kappa_l$.  Also, our calculated total $\kappa_l$  shows better agreement with Exp1 \cite{matsumoto2008systematic}  than the $\kappa_l$  calculated using the two-channel model by Wu et al. \cite{wu2019unusual} (see Supplementary Figure 4).

We note that we did not consider the effect the lattice thermal expansion in our calculations. In general, the thermal expansion will soften the phonon frequencies which give rise to the reduced phonon lifetimes and hence a lower-$\kappa_l$ at high temperatures. However, this reduction is partially counterbalanced by the increase in the $\kappa_l$ due to the anharmonic heat flux \cite{hardy1963energy} at high temperature, which is also absent in the current formalism. We did not calculate the thermal expansion coefficient and the anharmonic heat flux of TlInTe$_2$ as their calculations are computationally very expensive for non-cubic lattice. Nonetheless, the interplay of the thermal expansion and anhrmonic heat flux in TlInTe$_2$ and any ultralow-$\kappa_l$ materials in general is worth exploring in future studies.

In summary,  we have investigated the  microscopic origin and the underlying physical principles governing the extremely low and weakly temperature-dependent lattice thermal conductivity in the chain-like  crystalline semiconductor TlInTe$_2$ using a unified theory of lattice heat transport  that combines both the  particle-like  propagation  and wave-like tunneling of phonons. To treat the strong anharmonicity  present in TlInTe$_2$, we have applied the  SCPH theory to anharmonically renormalize the phonon frequencies at finite temperatures considering the quartic  anharmonicity. Our calculated $\kappa_l$ using the PBTE  coupled with  the SCPH  method (i.e., $\kappa_l$(SCPH)) and the off-diagonal contributions (i.e., $\kappa_l$(OD))   arising from the wave-like tunneling of phonons successfully reproduce  both the magnitude and temperature dependence (milder  than  T$^{-1}$ decay) of the measured  $\kappa_l$ in one set of experiments. Adding of  additional grain-boundary scatterings on top of  $\kappa_l$(SCPH)+ $\kappa_l$(OD), our calculated $\kappa_l$ reproduces well the second set of experiments. Our work thus highlights the important roles  of (i) temperature induced  renormalization of the harmonic phonon frequencies, particularly the low-energy rattling and other optical phonons, by the anharmonic effects, and (ii) the OD contributions  to $\kappa_l$ in the  heat-flux operator to correctly explain the origin of unusual lattice thermal transport of strongly anharmonic solids. The detailed microscopic understanding of the heat transfer mechanism obtained in this work will provide guidance towards the rational design and discovery of hitherto unknown ultralow-$\kappa_l$ compounds for various energy applications.

\section{Methods}

\subsection{Density functional theory calculations}
We perform all density functional theory (DFT) calculations using the Vienna Ab-initio Simulation Package (VASP) \cite{kresse1996efficiency, kresse1996efficient} employing the projector-augmented wave (PAW) \cite{kresse1999ultrasoft} potentials of 
Tl (5d$^{10}$ 6s$^{2}$ 6p$^{1}$), In (4d$^{10}$ 5s$^{2}$ 5p$^{1}$) and Te (5s$^{2}$ 5p$^{4}$) and utilized the
 PBEsol \cite{perdew2008restoring} parametrization of the generalized gradient approximation (GGA) \cite{perdew1996generalized} to the exchange-correlation energy functional.  We use a kinetic energy cut-off of 520 eV, and $\Gamma$-centered k-point mesh of 12 $\times$12$\times$12 to sample the Brillouin zone. The fully optimized lattice constants (a=8.406 $\AA$,  c=7.134 $\AA$) agree very well (absolute error $<$ 1  \%) with the experimentally reported values (a=8.478 $\AA$, c=7.185 $\AA$) \cite{banys1990powder}. We choose the 
 high-symmetry k-path in the Brillouin zone while potting the phonon dispersion following the convention of Setyawan et al. 
 \cite{setyawan2010high}.  Phonon dispersions are calculated using the second-order interatomic force constants (IFCs)  using  Phonopy \cite{phonopy}. Since the calculated phonon scattering rates strongly depends on the phonon frequencies  which can be quite sensitive to 
the supercell size, we   have performed convergence tests (see  Supplementary Figure 1), and adopted 2$\times$2$\times$2  supercell  for the calculations of the second-order IFCs.

\subsection{Thermal conductivity calculations}
To renormalize the phonon frequencies at finite temperature, and to calculate the lattice thermal conductivity ($\kappa_l$) using the PBTE, an accurate estimation of anharmonic IFCs,  namely the third-order and fourth-order IFCs  are required.   We obtain these anharmonic IFCs using the compressive sensing lattice dynamics (CSLD) method \cite{zhou2014lattice, tadano2015self, zhou2019compressive1, zhou2019compressive2} using  2$\times$2$\times$2  supercell and 6$\times$6$\times$6 $\Gamma$-centered k-point mesh.  While constructing the third and fourth order IFCs, a cut-off radius (r$_c$) is chosen, above which all three-body and four-body atomic interactions are discarded, respectively.  Although  with the increasing order of the IFCs, the atomic interactions become very short ranged, the value of r$_c$ can be  quite critical in correctly calculating $\kappa_l$ \cite{li2015orbitally}. We have performed  convergence tests of $\kappa_l$ against r$_c$ (see  Supplementary Figure 3) which are chosen to be 7.56 $\AA$ and 
 4 $\AA$ for the third and fourth-order IFCs, respectively. The r$_c$ value for the third-order IFCs is carefully  examined to give good convergence in  the calculated $\kappa_l$. While we have calculated the particle-like contributions to $\kappa_l$ using the PBTE considering the three-phonon, isotope and grain-boundary scatterings (see  Supplementary Note 2 for details), the off-diagonal contributions to $\kappa_l$ has been calculated by explicitly evaluating the off-diagonal terms of the heat-flux operator (see Supplementary Note 3 for details). Both these terms have been calculated by utilizing the anharmonically renormalized phonon frequencies obtained at finite temperatures. 

\subsection{Self-consistent phonon calculations}
 We solved the SCPH equations self-consistently until the phonon frequencies are converged to a given small threshold (e.g., 10$^{-3}$~cm$^{-1}$). using a relatively dense 6$\times$ 6$\times$6  mesh of $\mathbf{q}$-points. The renormalized phonon frequencies and eigenvectors are then used to obtain the renormalized harmonic IFCs through the inverse Fourier transformation. These renormalized IFCs are utilized to calculate phonon dispersions at finite temperatures.  We note that in this study we did not take into account the second-order correction to  $\omega_{\lambda}$ due to  the cubic anharmonic term because their contributions have been found less significant than the phonon frequency hardening by the quartic anharmonicity in some of the low-$\kappa_l$ systems such as PbTe \cite{xia2018revisiting} and clathrates \cite{tadano2018quartic}. However, we note that there are complex compounds like the tetrahedrites \ where the  role of the cubic anharmonicity is found to be quite  significant \cite{xia2020microscopic}.
We calculate the particle-like contributions, $\kappa_l$(PBTE), by iteratively solving the PBTE using the ShengBTE code \cite{li2014shengbte}. We use 12$\times$ 12$\times$12 $\mathbf{q}$-point mesh to obtain $\kappa_l$ with good convergence (see Figs. S3c-d). We present two components of the $\kappa_l$ in the results section: $\kappa_l^{\Vert}$ and  $\kappa_l^{\bot}$, which are parallel and perpendicular to the InTe$_4$ chain direction (i.e., along the c-axis in Fig. 1a) in the crystal structure of TlInTe$_2$, respectively. While $\kappa_l^{\Vert}$ is the zz-component of the $\kappa_l$-tensor, $\kappa_l^{\bot}$ is the directional average of the xx and yy-components of the $\kappa_l$-tensor. The average $\kappa_l$ is defined as the directional average of xx, yy and zz components of the $\kappa_l$-tensor throughout the manuscript. The T$^{-\alpha}$ fitting of the experimentally measured and calculated $\kappa_l$ data of TlInTe$_2$ are shown in  Supplementary Figure 5.

\section{Acknowledgements} 
\begin{acknowledgements}
We acknowledge financial supports from  the Department of Energy, Office of Science, Basic Energy Sciences
under grant DE-SC0014520 (thermal conductivity calculations), and  the U.S. Department of Commerce and National 
Institute of Standards and Technology as part of the Center for Hierarchical Materials Design (CHiMaD) under award 
no. 70NANB14H012 (DFT calculations). One of  us (Y.X.) is partially supported by the Toyota Research
Institute (TRI) through the Accelerated Materials Design and Discovery program (theory of anharmonic phonons). K.P. thanks Shashwat Anand for constructive comments on the manuscript. This work used computing resources provided by the (a) National Energy Research  Scientific Computing Center (NERSC),
 a U.S.  Department of Energy Office  of Science User Facility operated under  Contract No. DE-AC02-05CH11231, (b)  the Extreme Science and Engineering Discovery Environment (National Science Foundation Contract ACI-1548562), and (c) Quest high-performance computing facility at Northwestern University  which is jointly supported by the Office of the Provost,  the Office for  Research, and Northwestern University  Information Technology.
\end{acknowledgements}

\bibliography{tlinte2-ref}

\clearpage
\newpage

\end{document}